\documentclass{parcfd}

\usepackage{nopageno}
\usepackage{graphicx}
\usepackage{amsmath}
\usepackage{amsfonts}
\usepackage{amssymb}
\usepackage{hyperref}
\usepackage{layout}
\usepackage{algpseudocode}
\usepackage{algorithm}
\usepackage{caption}
\usepackage{subcaption}

\title{Monitoring the development of CFD applications on unstable HPC platforms}

\author{Damien Dosimont AND Guillaume Houzeaux}

\heading{Damien Dosimont and Guillaume Houzeaux}

\address{$^{*}$ Barcelona Supercomputing Center\\
Computer Applications in Science and Engineering (CASE)\\
Barcelona, Spain\\
e-mail: name.surname@bsc.es\\
web page: https://www.bsc.es/discover-bsc/organisation/scientific-structure/case
}


\abstract{We tackle the challenging tasks of monitoring on unstable HPC platforms the performance of CFD applications all along their development. 
We have designed and implemented a monitoring framework, integrated at the end of a CI-CD pipeline. 
Measures retrieved during the automatic execution of production simulations are analyzed within a visual analytics interface we developed, providing advanced visualizations and interaction.
We have validated this approach by monitoring the CFD code Alya over two years, detecting and resolving issues related to the platform, and highlighting performance improvement.}
\begin{document}

\section{INTRODUCTION}

Monitoring the performance of a CFD code throughout its development on the HPC production platform has become a requirement to ensure its stability. CI-CD pipelines are ideal for conducting this task. For instance, Sampedro and al.~\cite{sampedro_continuous_2018} use Jenkins and Singularity to perform HPC performance monitoring on small-scale examples. 

However, the lack of stability of the platform behavior can introduce noise in the performance analysis.
Overheating,  file system misusing, or bandwidth abuse can heavily impact the behavior of the applications running on the affected platform. 
To tackle this issue, Voss and al.~\cite{voss_automated_2017} have designed a system health and performance monitoring tool that enables tracking both health and historical performance by running automatically different benchmark suites. 
Larrea and al.~\cite{larrea_use_2015} have carried out a comparison between different CI-CD frameworks and their abilities to manage performance monitoring on HPC platforms using various types of benchmarks at a large scale.
Although their study helped find inconsistencies in their HPC platform, they deplore the limited plotting capabilities and data management of the CI-CD tools they evaluated.

Our proposal is a performance monitoring framework that fits in a CI-CD pipeline.
It aims to ensure the performance stability of the application, taking into account the history of the passed runs to reduce the noise created by the HPC platform on the analysis. 
A visual analytics interface offers the developers interactions and various visualizations to explore advanced metrics retrieved during the monitoring campaign.
We have validated this method with BSC's in-house code \emph{Alya}~\cite{vazquez_alya_2016} over two years on simulations running with up to 768 MPI processes, detecting punctual and persistent problematic behaviors of the HPC platform Marenostrum that we could help resolve and highlighting performance improvements related to various code enhancements.

\section{MONITORING THE DEVELOPMENT OF CFD APPLICATIONS}

Our method takes place at the last part of the CI-CD workflow. 
As the development of \emph{Alya}~\cite{vazquez_alya_2016}, is centralized within Gitlab, we can benefit from its CI-CD pipeline features, but our strategy can be adapted to other CI-CD tools like Jenkins or Travis CI. 

Slight instrumentation of the code is necessary to get measures that will feed the analysis, such as phase duration, memory usage, or parallelism efficiency. The application computes those measures but follows defined data structure specifications. Since the data is asemantic, whatever kind of metrics can be exported. Our data structure also enables the developer to define a hierarchy between the measures, which is useful for representing the different levels of an execution stack. We also assign labels to the measures, for instance, the type of operations (IO, computation, communications). 


We have developed a framework, \emph{alya-cicd}, written in python, which is responsible for 1) building the CFD application on the production platform with all the compilers and flavors desired by the developers, 2) running different simulations on the production platform, 3) retrieving the performance measures from the simulation executions, 4) formatting the data, adding metadata like the job information, the environment variables, the build parameters, etc. to get the maximum information on the execution context and storing them in a database that contains all the passed simulations. 
The execution of \emph{alya-cicd} is automatically triggered and monitored by the CI-CD pipeline that notifies the developers of the end of the simulation campaign.

Finally, the developers analyze the data through a visual analytics framework, \emph{rooster}, connected to the database in which are stored the results of all the passed executions. 
Several levels of visualizations are available, hierarchized depending on the level of detail they provide, and interconnected through user interaction. 
Showing the source code differences between two successive commits clotures the top-down analysis.

%

Rooster is developed in Python and relies on Dash and Plotly libraries. It is deployed as a server that the developers can consult at any moment.

\section{RESULTS}

\begin{figure}
\centering
\begin{minipage}[b]{.4\textwidth}
    \centering
    \includegraphics[width=.8\textwidth]{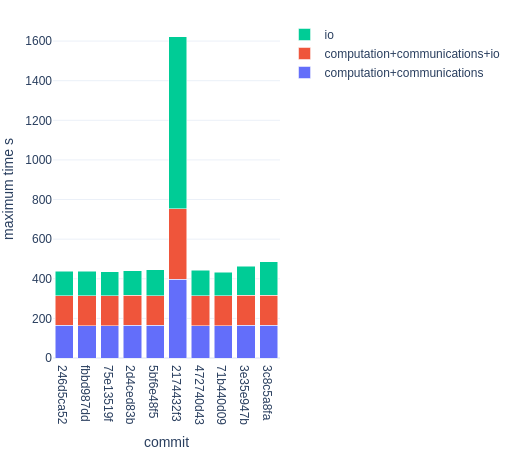}
    \caption{Stack bar chart showing the time passed in the different operation types. The 6$^{th}$ simulation iteration is affected by longer IO}
    \label{fig:gpfs}
\end{minipage}%
\hspace{0.1\textwidth}%
\begin{minipage}[b]{.4\textwidth}
    \centering
    \includegraphics[width=\textwidth]{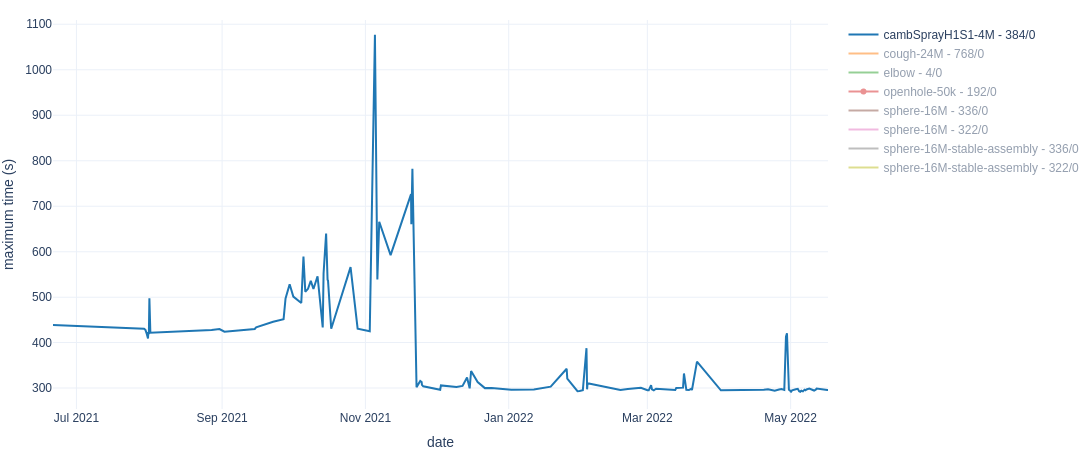}
    \caption{Simulation time of CambSprayH1S1, decreasing after November 2021}
    \label{fig:cambSpray}
\end{minipage}
\end{figure}

\paragraph{GPFS collapse}
Some unaware users sometimes perform heavy IO operations on Marenostrum's GPFS, collapsing more or less critically the file system and degrading the performance of other users' applications. 
This situation has been highlighted several times by our framework. 
Figure~\ref{fig:gpfs} shows how labeling the type of operations during the instrumentation (computation, communications, or IO) helps to highlight that the increasing execution duration of one of the simulations is mainly due to IO operations, confirming the file system collapse. 
The next executions show the performance goes back to normal.

\paragraph{MPIO hints}
Alya's IO operations are realized through MPI-IO to ensure full parallelism of the whole data flow. 
On Marenostrum IV, the configuration of ROMIO, the implementation of MPI-IO, was not set properly, and the MPI-IO operations were not optimized.
After discovering the IO operations were abnormally long in the CambSprayH1S1 simulation~\cite{both_assessment_2022}, which is IO intensive, we asked the support team to improve the configuration of ROMIO. 
The improvement after correction is visible in Figure~\ref{fig:cambSpray}, at the end of November 2021.

\begin{figure}
\centering
\begin{subfigure}{.5\textwidth}
  \centering
  \includegraphics[width=\linewidth]{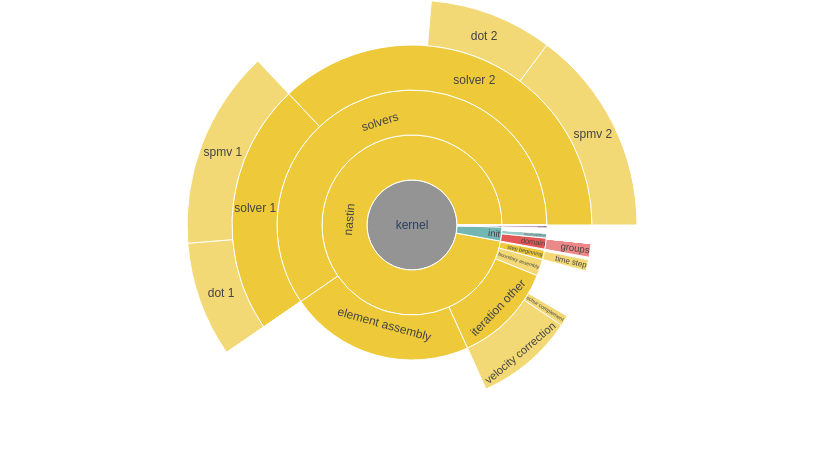}
  \caption{Before vectorization}
  \label{fig:sub1}
\end{subfigure}%
\begin{subfigure}{.5\textwidth}
  \centering
  \includegraphics[width=\linewidth]{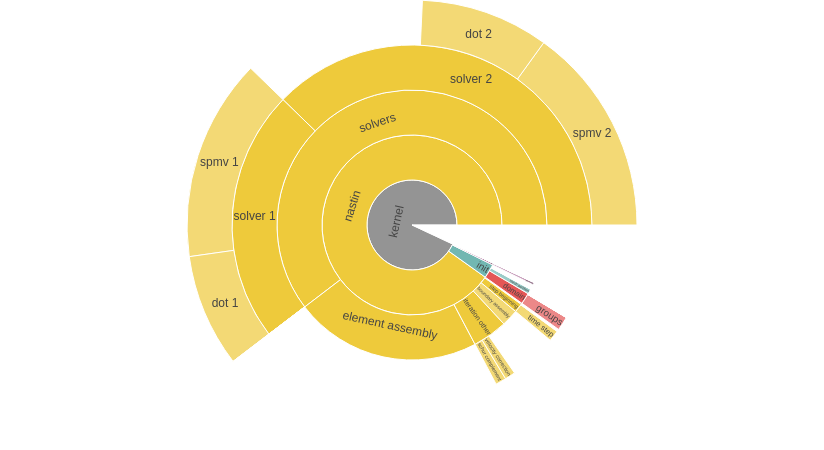}
  \caption{After vectorization}
  \label{fig:sub2}
\end{subfigure}
\caption{Sunbursts of the cough simulation showing the execution time improvement thanks to the vectorization of the velocity correction}
\label{fig:cough}
\end{figure}

\paragraph{Optimization of the Cough simulation}
The Cough use case~\cite{calmet_large_2021} uses Alya's Navier-Stoke fluid mechanic module.
One of the loops of the subroutine \texttt{nsi\_velocity\_correction} (previously \texttt{nsi\_elmcor}) has been vectorized commit \texttt{3c8e81f6}\footnote{https://alya.gitlab.bsc.es/alya/open-alya/-/commit/62b135018}. 
The performance improvement is visible in Rooster's timelines. 
The sunburst visualizations Figure~\ref{fig:cough} show the reduction of the velocity correction duration before and after the modification.

\section{CONCLUSION AND FUTURE WORKS}

We have presented a framework to monitor the development of CFD applications, supported by the example of our in-home CFD code, Alya, but extensible to any HPC code. This framework aims at tracking the performance variation of the application on the HPC production platform. It helps eliminate the noise induced by the system thanks to the representation of the execution history, the characterization of the different types of operations, and a visual analytics approach providing a top-down analysis flow ending with the comparison between the source codes of successive commits.
Our future work will consist in putting more emphasis on getting knowledge on the infrastructure behavior.  
We would like to retrieve hardware counters during the execution (e.g., bandwidth usage, and temperature) to correlate them to the application behavior.
We are also interested in running periodically stable benchmarks (computation, communication, or IO-intensive) to help identify the instability on the platform.
We would like to let the analysts tag and describe platform issues and code significant changes as a form of events that will be integrated into the timeline.
Last but not least, we will add new simulations of bigger sizes to test the scalability of our framework.

%

\bibliographystyle{ieeetr}
\small
\bibliography{parcfd2023}

\end{document}